\documentclass[prd, preprint, longbibliography, 12pt]{revtex4-1}

\usepackage{commath}
\usepackage{amsmath}
\usepackage{amssymb}
\usepackage{setspace}
\usepackage{graphicx}
\usepackage{natbib}
\usepackage{float}
\usepackage{tensor}
\usepackage[utf8]{inputenc}
\usepackage{amsfonts}
\usepackage{braket}
\usepackage{esint}
\usepackage{breqn}
\usepackage{IEEEtrantools}
\usepackage{hyperref}

\usepackage[english]{babel}
\usepackage{graphicx}
\usepackage{amsmath}

\usepackage{amssymb}
\usepackage{multirow}
\usepackage{url}
\usepackage{tikz}
\usepackage{braket}
\usetikzlibrary{decorations.markings}
\usepackage{subcaption}
\usepackage{cancel}
\usepackage{mathtools}
\usepackage[utf8]{inputenc}
\usepackage{xargs, ifthen}
\usepackage[breakwords]{truncate}

\usepackage{hyperref}
\DeclareMathAlphabet{\pazocal}{OMS}{zplm}{m}{n}

\allowdisplaybreaks

\def\s[#1\s]{\begin{align}\begin{split}#1\end{split}\end{align}}
\def\[#1\]{\begin{align}#1\end{align}}

\begin{document}
 
 %

\begin{center}
{ \large \bf Octonions, trace dynamics, and non-commutative geometry}\\
{\large - {\it A case for unification in spontaneous quantum gravity} - }

\vskip 0.2 in

{\large{\bf Tejinder P.  Singh }}

\medskip

{\it Tata Institute of Fundamental Research, Homi Bhabha Road, Mumbai 400005, India}\\

\bigskip

 \; {\tt tpsingh@tifr.res.in}

\end{center}

\bigskip

\centerline{\bf ABSTRACT}
\noindent  We have  recently proposed a new matrix dynamics at the Planck scale, building on the theory of trace dynamics and Connes non-commutative geometry programme. This is a Lagrangian dynamics in which the matrix degrees of freedom are made from Grassmann numbers, and the Lagrangian is trace of a matrix polynomial. Matrices made from even grade elements of the Grassmann algebra are called bosonic, and those made from odd grade elements are called fermionic: together they describe an `aikyon'. The Lagrangian of the theory is invariant under global unitary transformations, and describes gravity and Yang-Mills fields coupled to  fermions. In the present article we provide a basic definition of spin angular momentum in this matrix dynamics, and introduce a  bosonic(fermionic) configuration variable conjugate to the spin of a boson(fermion). We then show that at energies below Planck scale, where the matrix dynamics reduces to quantum theory, fermions have half-integer spin (in multiples of Planck's constant), and bosons have integral spin. We also show that this definition of spin agrees with the conventional understanding of spin in relativistic quantum mechanics. Consequently, we obtain an elementary proof for the spin-statistics connection.
We then motivate why an octonionic space is the natural space in which an aikyon evolves.
The group of automorphisms in this space is the exceptional Lie group $G_2$ which has fourteen generators [could they stand for the twelve vector bosons and two degrees of freedom of the graviton ?]. The aikyon also resembles a closed string, and it has been suggested in the literature  that 10-D string theory can be represented as a 2-D string in the 8-D octonionic space. From the work of Cohl Furey and others it is known that
the Dixon algebra made from the four division algebras [real numbers, complex numbers, quaternions and octonions] can possibly describe  the symmetries of the standard model. In the present paper we outline how  in our work the Dixon algebra arises naturally, and could lead to a unification of gravity with the standard model. From this matrix dynamics, 
local quantum field theory arises as a low energy limit of this Planck scale dynamics of aikyons, and classical general relativity arises as a consequence of spontaneous localisation of a large number of entangled aikyons. We propose that classical curved space-time and Yang-Mills fields arise from an effective gauging which results from the collection of symmetry groups of the spontaneously localised fermions. Our work suggests that we live in an eight-dimensional octonionic universe: four of these dimensions constitute space-time and the other four constitute the octonionic  internal directions on which the standard model forces live.

\bigskip

\bigskip
\section{Introduction}
We have recently proposed a new matrix dynamics at the Planck scale \cite{maithresh2019, MPSingh}  building on Adler's theory of trace dynamics \cite{Adler:04, Adler:94, AdlerMillard:1996}  and by using constructs from Connes' non-commutative geometry programme 
\cite{Connes2000, Chams:1997}
to incorporate gravity and curvature into trace dynamics. One starts by assuming the existence of a Riemannian differentiable manifold along with the standard Dirac operator $D_B\equiv i\gamma^\mu\nabla_\mu$. Matter is described by relativistic point particles. Einstein field equations are {\it not} assumed. As is known from earlier work, the information about metric and curvature can also be captured by the Dirac operator and its eigenvalues \cite{Rovelli, Landi1999}. 

Given this classical background, the transition to the matrix dynamics is made as follows. The idea is to describe fundamental degrees of freedom by matrices, instead of by real numbers. The motivation is to achieve a formulation of quantum field theory which does not refer to classical time. Doing so also allows one to construct the new dynamics at the Planck scale, from which quantum field theory and classical general relativity are emergent at lower energies and at length and time scales much larger than Planck length and Planck time. Given a matter Lagrangian on a space-time background, all configuration variables and their corresponding velocities are replaced by matrices, and the trace of the resulting matrix polynomial defines the new Lagrangian. Integral of this Lagrangian over time defines the action, whose variation gives the matrix-valued Lagrange equations of motion. These equations define the Lagrangian dynamics, for which an equivalent Hamiltonian dynamics can also be constructed, following standard techniques \cite{Adler:04}.

The next step is to raise space-time points also to the status of matrices, and employ the Dirac operator to describe distance and curvature on the resulting non-commutative geometry. One no longer makes a distinction between the matrix describing a relativistic particle, and the matrix describing the space-time geometry it produces. Together, they are described by a Grassman-valued matrix $q$, which can always be written as a sum: $q\equiv q_B + q_F$, where the bosonic matrix $q_B$ is made of even-grade elements of the Grassmann algebra, and the fermionic matrix $q_F$ is made of odd-grade elements of the Grassmann algebra. $q_F$ describes the matter part, and $q_B$ describes the contribution of $q_F$ to space-time geometry and gravity. 
We call this entity an `atom' of space-time-matter, or an aikyon. It evolves in Hilbert space, with evolution described by a time-parameter $\tau$ intrinsic to a non-commutative geometry  \cite{Connes2000, Connes1994}, and labelled by us as Connes time. The matrix dynamics Lagrangian and action for the aikyon $q$ are given by \cite{maithresh2019}
\begin{equation}
 \frac{S}{C_0}  =  \frac{1}{2} \int \frac{d\tau}{\tau_{P}} \; Tr \bigg[\frac{L_P^2}{L^2c^2}\; \left(\dot{q}_B +\beta_1 \frac{L_P^2}{L^2}\dot{q}_F\right)\; \left(\dot{q}_B +\beta_2 \frac{L_P^2}{L^2} \dot{q}_F\right) \bigg]
 \label{acnaik}
\end{equation}
where $\beta_1$ and $\beta_2$ are  constant  fermionic numbers. These Grassmann numbers make the Lagrangian bosonic.
The only two fundamental constants are Planck length and Planck time - these scale the length scale $L$ of the aikyon, and the Connes time, respectively. $C_0$ is a constant with dimensions of action, which will be identified with Planck's constant in the emergent theory. The Lagrangian and action are not restricted to be self-adjoint. A dot denotes derivative with respect to Connes time
(${\rm Dot} = \delta/c\delta \tau$).  By varying this action w.r.t. $q_B$ and $q_F$ one gets a pair of coupled equations of motion, which can be solved to find the evolution of $q_B$ and $q_F$. The respective momenta $p_B$ and $p_F$ are constants of motion, and the expression for $p_B$ can be written as an eigenvalue equation for the modified Dirac operator $D\equiv D_B + D_F$:
\begin{equation}
\left[D_B + D_F\right] \psi =  \frac{1}{L} \bigg(1+ i \frac{L_P^2}{L^2}\bigg)\psi
\label{dfull}
\end{equation}
where
\begin{equation}
D_B \equiv \frac{1}{Lc}\;  \frac{dq_B}{d\tau}; \qquad   D_F \equiv \frac{L_P^2}{L^2}\frac{\beta_1+\beta_2}{2Lc}\;  \frac{dq_F}{d\tau}
\end{equation}
$D_B$ is defined such that in the commutative c-number limit where space-time emerges, it becomes the standard Dirac operator on a Riemannian manifold. 
If there are many aikyons in the theory, their total action is the sum of their individual actions. There is no classical space-time in this dynamics at the Planck scale; only a Hilbert space, and a soon to be introduced octonionic space, from which space-time is emergent.

We next asked what the coarse-grained matrix dynamics looks like, at energies much lower than Planck scale; equivalently, at times scales much larger than Planck time. This question can be answered by employing the techniques of statistical thermodynamics, as set up in the theory of trace dynamics. The low energy dynamics falls in two classes. If not too many aikyons are entangled with each other, the anti-self-adjoint component of the net Hamiltonian is negligible, and the emergent dynamics is quantum dynamics without a classical space-time. The canonical variables obey quantum commutation relations, and the Heisenberg equations of motion, for which there is also an equivalent Schr\"{o}dinger picture. Evolution is still in Connes time, and there is no background space-time, yet.

The other limiting class is when sufficiently many aikyons get entangled; then the anti-self-adjoint part of the net Hamiltonian becomes significant. This causes rapid spontaneous localisation, loss of quantum superposition, and the emergence of classicality. The classical space-time manifold emerges, and its points are defined by the position eigenvalues to which the fermions localise. The metric and classical curvature are  recovered by localisation of the Dirac operators of the aikyons to their specific eigenvalues. The net action for the aikyons described above reduces to the action for classical general relativity. In this way Einstein field equations are recovered, with relativistic point particles as sources. Given this space-time background, the above-mentioned quantum dynamics of uncollapsed aikyons can be described as quantum field theory on a background space-tiime. This background is generated by the classicalised matter degrees of freedom.

Subsequently, we have generalised this action to include Yang-Mills gauge fields \cite{MPSingh}, and the new action is given by $S/C_0=\int \tau\;{\cal L} / \tau_{P}$ where
\s[
\mathcal{L} = Tr \biggl[\biggr. \dfrac{L_{p}^{2}}{L^{4}} \biggl\{\biggr. i\alpha \biggl(\biggr. q_{B} + \dfrac{L_{p}^{2}}{L^{2}}\beta_{1} q_{F} \biggl.\biggr) + L \biggl(\biggr. \dot{q}_{B} + \dfrac{L_{p}^{2}}{L^{2}}\beta_{1}\dot{q}_{F} \biggl.\biggr) \biggl.\biggr\}\\
\biggl\{\biggr. i\alpha \biggl(\biggr. q_{B} + \dfrac{L_{p}^{2}}{L^{2}}\beta_{2} q_{F} \biggl.\biggr) + L \biggl(\biggr. \dot{q}_{B} + \dfrac{L_{p}^{2}}{L^{2}}\beta_{2} \dot{q}_{F} \biggl.\biggr) \biggl.\biggr\} \biggl.\biggr]
\label{laym}
\s]
This Lagrangian for an aikyon should be compared with the earlier one (\ref{acnaik}) which had only gravity and Dirac fermions as unified components of the aikyon. This new Lagrangian here also includes gauge-fields and their currents, through $q_B$ and $q_F$, assumed self-adjoint. $\alpha$ is the Yang-Mills coupling constant, assumed to be a real number.   Gravitation, and Yang-Mills fields, and their corresponding sources,  are unified here as the `position' $q$ and `velocity' $dq/d\tau$ of the aikyon. With position being the Yang-Mills part, and velocity being the gravitation part. 

By defining new dynamical variables $\dot{\widetilde Q}_B$ and $\dot{\widetilde{Q}}_F$ as
\begin{equation}
{\dot{\widetilde{Q}}_B} = \frac{1}{L} (i\alpha q_B + L \dot{q}_B); \qquad  {\dot{\widetilde{Q}}_F} = \frac{1}{L} (i\alpha q_F + L \dot{q}_F)
\label{lilqu}
\end{equation}
this Lagrangian can be brought to the elegant and revealing form
\[
 \mathcal{L} = Tr \biggl[\biggr. \dfrac{L_{p}^{2}}{L^{2}} \biggl(\biggr. \dot{\widetilde{Q}}_{B} + \dfrac{L_{p}^{2}}{L^{2}} \beta_{1} \dot{\widetilde{Q}}_{F} \biggl.\biggr) \biggl(\biggr. \dot{\widetilde{Q}}_{B} + \dfrac{L_{p}^{2}}{L^{2}} \beta_{2} \dot{\widetilde{Q}}_{F} \biggl.\biggr) \biggl.\biggr]
\label{eq:tracelagn}
\]
We used this form to express a unification for gravity and gauge fields in our recent work \cite{MPSingh} in terms of these new complex variables. These variables incorporate the position and velocity of the aikyon as their real and imaginary parts. We note that $q_B, q_F, \dot{q}_B$ and $\dot{q}_F$ are all assumed to be self-adjoint in the present paper. This new construction amounts to re-expressing the aikyon $q$ by the variable $\widetilde{Q}\equiv \widetilde{Q}_B + \widetilde{Q}_F$ where the bosonic $\widetilde{Q}_B$ and the fermionic $\widetilde{Q}_F$ are further expressed in terms of their self-adjoint and anti-self-adjoint parts as in Eqn. (\ref{lilqu}) above. The self-adjoint part is velocity, which encodes gravity, and the anti-self-adjoint is position, which encodes Yang-Mills gauge fields. This natural split of a Grassmann matrix into its four parts [bosonic self-adjoint, bosonic antii-self-adjoint, fermionic self-adjoint, fermionic anti-self-adjoint] captures gravity and Yang-Mills fields, as well as their sources. There is only one term in the Lagrangian of an aikyon, which when opened up using this split, gives rise to sixteen different terms. The classical limit is Einstein gravity coupled to Yang-Mills 
fields and matter sources \cite{MPSingh}. 

Our new Planck scale matrix dynamics has been used to make several predictions. We derived the Bekenstein-Hawking black hole entropy from the microstates of its constituent aikyons \cite{maithresh2019b}. We have predicted the Karolyhazy uncertainty relation as a consequence of our theory \cite{Singh:KL}. We have used the theory to propose that dark energy is a large-scale quantum gravitational phenomenon \cite{Singh:DE}. We have explained the remarkable fact that the Kerr-Newman black hole has the same gyromagnetic ratio as a Dirac fermion, both being twice the classical value \cite{MPSingh}. 

In the present article we provide a basic definition of spin angular momentum in this matrix dynamics, and introduce a  bosonic(fermionic) configuration variable conjugate to the spin of a boson(fermion). We then show that at energies below Planck scale, where the matrix dynamics reduces to quantum theory, fermions have half-integer spin (in multiples of Planck's constant), and bosons have integral spin. We also show that this definition of spin coincides with the conventional understanding of spin in relativistic quantum mechanics. Consequently, we obtain an elementary proof for the spin-statistics connection. Essentially, we reverse the arguments of the traditional proof of spin-statistics connection in relativistic quantum field theory \cite{streater}. Instead of showing that integer-spin particles obey Bose-Einstein statistics, we show that particles obeying Bose-Einstein statistics have integer spin. Similarly, we show that particles obeying Fermi-Dirac statistics have half-integer spin.

We then motivate why an octonionic space is the natural space in which an aikyon evolves.
The group of automorphisms in this space is the exceptional Lie group $G_2$ which has fourteen generators [could they stand for the twelve vector bosons and two degrees of freedom of the graviton ?]. The aikyon also resembles a closed string, and it has been suggested in the literature  that 10-D string theory can be represented as a 2-D string in the 8-D octonionic space. From the work of Cohl Furey and others it is known that
the Dixon algebra made from the four division algebras [real numbers, complex numbers, quaternions and octonions] can possibly describe  the symmetries of the standard model. In the present paper we outline how  in our work the Dixon algebra arises naturally, and could lead to a unification of gravity with the standard model. From this matrix dynamics, 
local quantum field theory arises as a low energy limit of this Planck scale dynamics of aikyons, and classical general relativity arises as a consequence of spontaneous localisation of a large number of entangled aikyons. We propose that classical curved space-time and Yang-Mills fields arise from an effective gauging which results from the collection of symmetry groups of the spontaneously localised fermions. Our work suggests that we live in an eight-dimensional octonionic universe: four of these dimensions constitute space-time and the other four constitute the octonionic  internal directions on which the standard model forces live.

\section{A definition for spin in the new matrix dynamics}
Our starting point is the Lagrangian for an aikyon, as given in Eqn. (62) of \cite{MPSingh}, and mentioned above in (\ref{eq:tracelagn}), which we reproduce here again:
\[
 \mathcal{L} = Tr \biggl[\biggr. \dfrac{L_{p}^{2}}{L^{2}} \biggl(\biggr. \dot{\widetilde{Q}}_{B} + \dfrac{L_{p}^{2}}{L^{2}} \beta_{1} \dot{\widetilde{Q}}_{F} \biggl.\biggr) \biggl(\biggr. \dot{\widetilde{Q}}_{B} + \dfrac{L_{p}^{2}}{L^{2}} \beta_{2} \dot{\widetilde{Q}}_{F} \biggl.\biggr) \biggl.\biggr]
\label{eq:tracelag}
\]
We now introduce self-adjoint bosonic operators $R_B$ and $\theta_B$, and self-adjoint fermionic operators $R_F$ and $\theta_F$, as follows:
\begin{equation}
\widetilde{Q}_B \equiv R_B\; \exp i\theta_B \ ;   \qquad \widetilde{Q}_F \equiv R_F \; \exp i\eta \theta_F
\label{que}
\end{equation}
Here, $\eta$ is a real Grassmann number, introduced to ensure that the fermionic phase is bosonic, so that $\widetilde{Q}_F$ comes out fermionic, as desired, upon the Taylor expansion of its phase. As is known, this will give that $\exp i\eta\theta_F = 1+ i\eta\theta_F$ with the higher terms in the Taylor expansion vanishing, because $\eta^2=0$. These definitions are equivalent to expressing a Grassmann-valued matrix in terms of its `amplitude' matrix and `phase' matrix, as if to represent the matrix on a complex plane.

We note from the definition of $\widetilde{Q}_B$ that  it  remains unchanged  under the shift $\theta_B \rightarrow \theta_B + 2\pi I$. In this sense $\theta_B$ acts like  an angle variable, and we will require all bosonic physical quantities dependent on $\theta_B$ to remain unchanged under the shift $\theta_B \rightarrow \theta_B + 2\pi I$. We can also reason why the fermionic  $\widetilde{Q}_F$ should change sign if the bosonic part of the corresponding aikyon undergoes a shift  $\theta_B \rightarrow \theta_B + 2\pi I$. The bosonic $\widetilde{Q}_B$ is a matrix made from  elements of the even-grade Grassmann algebra, so that in principle we can consider the case that it is made from two fermionic matrices:
$\widetilde{Q}_B= \widetilde{Q}_{F1} \times \widetilde{Q}_{F2}$. Hence
\begin{equation}
\widetilde{Q}_B \equiv R_B\; \exp i\theta_B  =   \widetilde{Q}_{F1} \times \widetilde{Q}_{F2} = R_{F1} \; \exp i\eta \theta_{F1}
\times R_{F2} \; \exp i\eta \theta_{F2}
\label{que2}
\end{equation}
The shift $\eta\theta_{F1} \rightarrow \eta\theta_{F1}+\pi$ induces a sign change in $\widetilde{Q}_{F1}$ and a simultaneous 
shift $\eta\theta_{F2} \rightarrow \eta\theta_{F2}+\pi$ induces a sign change in $\widetilde{Q}_{F2}$. Together these two sign changes imply that the bosonic $\widetilde{Q}_{B}$ does not change sign, and one can conclude that these two shifts are equivalent to 
$\theta_B \rightarrow \theta_B + 2\pi I$. Conversely, under $\theta_B \rightarrow \theta_B + 2\pi I$, the fermionic parts each undergo a change of sign. They do not change under $\theta_B \rightarrow \theta_B + 4\pi I$. This observation is analogous to the fact that a spinor changes sign under a $2\pi$ rotation in space. We may think of spinors as eigenstates of fermionic matrices, having odd-grade Grassmann numbers as their components. Vectors are eigenstates of bosonic matrices, and clearly, a product of two spinors is a vector, just as a product of two fermionic matrices is a bosonic matrix.

Each of these four newly introduced  self-adjoint operators are functions of Connes time, and are the four configuration variables which define the aikyon. By substituting these definitions of $\widetilde{Q}_B$ and $\widetilde{Q}_F$ in the above Lagrangian (\ref{eq:tracelag}), we can write the Lagrangian in terms of time derivatives of these four configuration variables. We will do that in the next section. For now, it suffices to note that the canonical linear momenta $p_{BR}$ and $p_{FR}$ are defined as usual, as derivatives of the Lagrangian with respect to the corresponding velocities:
\begin{equation}
p_{BR} = \frac{\delta {\mathcal L}} {\delta \dot {{R}}_{B}} \ ; \qquad p_{FR} = \frac{\delta {\mathcal L}} {\delta \dot {{R}}_{F}}
\end{equation}
The novel part is the following. We define bosonic and fermionic {\it spin angular momenta} as follows:
\begin{equation}
p_{B\theta } = \frac{\delta {\mathcal L}} {\delta \dot {{\theta}}_{B}} \ ; \qquad p_{F\theta } = \frac{\delta {\mathcal L}} {\delta \dot {{\theta}}_{F}}
\end{equation}
[A word about dimensions. The Lagrangian and action as introduced  here are dimensionless, and hence so is the linear momentum. However, when care is taken of the $\tau_P$ present in the action integral, and the $C_0$ on the left hand side of the action integral brought to the right, linear momentum acquires familiar correct dimensions. The same reasoning applies for the dimensions of angular momentum]. 
The following proof for spin quantisation is independent of the specific form of the Lagrangian for the matrix dynamics. All that is required is that the configuration variables have a self-adjoint part as well as an anti-self-adjoint part.
As is known from Adler's theory of trace dynamics, and is true also for the present matrix dynamics, there is a conserved charge known as the Adler-Millard charge \cite{RMP:2012}. This charge results from the invariance of the trace Lagrangian under global unitary transformations of the degrees of freedom. The charge has dimensions of action and is denoted by the symbol $\tilde{C}$:
\begin{equation}
    \tilde{C} = \sum_{r\in B}[q_r,p_r] -\sum_{r\in F} \{q_r,p_r \} 
    \label{amc}
\end{equation}
which is the sum over the shown commutators for bosonic degrees of freedom, minus the sum over the shown anti-commutators  for fermionic degrees of freedom. If there are many aikyons in the system, the conserved charge is the sum over all aikyons, of their individual contributions. For the present set of momenta, the Adler-Millard charge is
\begin{equation}
\tilde{C} = [R_B, p_{BR}] + [\theta_B, p_{B\theta }] - \{R_F, p_{FR}\} - \{\theta_F, p_{F\theta }\}
\end{equation}

As we know from trace dynamics and our own earlier work, if we observe this matrix dynamics at energy scales much lower than Planck scale, the emergent dynamics is quantum theory. This is shown by coarse-graining the matrix dynamics over times much larger than Planck times, and using the techniques of statistical thermodynamics to find out the coarse-grained dynamics  [There is an additional requirement that any anti-self-adjoint component in the momenta and in the Hamiltonian must be negligible, for the emergence of quantum theory].  In particular, the Adler-Millard charge gets equipartitioned over all the degrees of freedom, and the constant value of the equipartitioned charge per degree of freedom is identified with Planck's constant $\hbar$. This implies, from the structure of the charge $\tilde{C}$ above,  that the self-adjoint part of statistically averaged canonical variables (identified with the dynamical variables of quantum field theory) obey the canonical commutation relations of quantum theory:
 \begin{equation}
[R_B, p_{BR}] =i\hbar\ ; \qquad [\theta_B, p_{B\theta }] = i\hbar \ ; \qquad \{R_F, p_{FR}\} = i\hbar \ ; \qquad \{\theta_F, p_{F\theta }\} =i\hbar
\end{equation}
It is understood in these commutators that only the self-adjoint component of the momenta is present, and this component has been averaged over the canonical ensemble of the microstates allowed at statistical equilibrium.
From here, it is possible to deduce the quantisation of spin angular momentum. From the second commutation relation - between $\theta_B$ and $p_{B\theta }$, we deduce that this spin angular momentum is a displacement operator, whose eigenvalues are quantised:
\begin{equation}
p_{\theta B} = - i \hbar \frac{\delta \ }{ \delta \theta_B} ;  \quad  - i \hbar \frac{\delta \ }{ \delta \theta_B}\psi = \lambda \psi; \quad 
\psi \sim \exp \left[ i \frac{\lambda}{\hbar} \theta_B \right] \implies \lambda = n \hbar
\end{equation}
where $n$ is an integer. Moreover, since $\theta_B$ is an even grade Grassmann matrix, two such matrices commute, leaving the state of a multi-particle bosonic system unchanged upon interchange of two identical bosons. The state is hence symmetric, and the system obeys Bose-Einstein statistics.

The situation regarding fermions is more subtle. Because the fermionic spin $p_{F\theta }$ satisfies an {\it anti-commutation} relation with the dynamical variable $\theta_F$, one can construct a displacement operator for it using Berezin calculus. [By itself,  $\theta_F$ does not permit any angle interpretation for itself. However we can infer fermion spin quantisation indirectly. Consider a bosonic degree of freedom $B$ made from a product of two identical fermions $F_1$ and $F_2$, i.e. $B=F_1 F_2$. Since $B$ has integral spin and since spin is additive, and since we cannot discriminate between the contribution of spin from $F_1$ and spin from $F_2$, we can conclude the following. If the boson has spin $\hbar$, the fermions each have spin $\hbar/2$. And because fermions are made from odd-grade Grassmann numbers which anti-commute, a state for a system of identical fermions is anti-symmetric under exchange of particles, implying that the statistics is Fermi-Dirac].

The fermionic Berezin displacement operator corresponding to $\theta_F$ is $i\hbar \delta /\delta \theta_F$ and its eigenvalue will be a Grassmann number, not a c-number. However, we can construct the bosonic operator $i\hbar \delta /\delta \theta'_F$ where $\theta'_F = \eta\theta_F$ and this has a c-number eigenvalue $\lambda$ and an eigenstate proportional to $\exp [-i\lambda\eta\theta_F/\hbar]$. Since the phase must change sign under a shift of $2\pi$ in $\eta\theta_F$, it follows that the fermion has a spin $\lambda = \pm\hbar/2$. 

The novelty of the present proposal is the introduction of the fermionic configuration variable $\theta_F$. There is no analogue for it in quantum mechanics. That is because one develops quantum mechanics by quantising classical dynamical theories. In so doing, we never arrive at  this dynamical variable $\theta_F$, which indeed comes down to us from the Planck scale matrix dynamics. Moreover, there is no space-time, yet, in our analysis. This is another piece of evidence to suggest that quantum mechanics is a low energy limit of a [more complete] underlying dynamics: a dynamics in which classical space-time is absent. It also seems to be the case that this proof of the spin-statistics connection does not manifestly require a space-time symmetry such as  Lorentz invariance.

Next, we show, using our specific Lagrangian,  that the spin angular momentum introduced here agrees with the conventional understanding of spin in relativistic quantum mechanics.

\section{Relating the spin in matrix dynamics to the spin in quantum mechanics}
We work out the expressions for the four momenta by first substituting the forms (\ref{que}) into the Lagrangian (\ref{eq:tracelagn}). The velocities are given by (assuming the small angle approximation)
\begin{equation}
\dot{\widetilde Q}_B = \dot{R}_B \; \exp i\theta_B + R_B \exp i\theta_B \times i \dot{\theta}_B   \ ; \quad \dot{\widetilde{Q}}_F = \dot{R}_F \exp i\eta\theta_F  + R_F \exp i\eta\theta_F \times i\eta\dot{\theta}_F 
\label{qexp2}
\end{equation}
These are substituted in the Lagrangian, and they yield the following expressions for the four momenta. We first open the brackets in the expression for the Lagrangian, and write it as a sum of four terms: ${\cal L} = T_1 + T_2 + T_3 + T_4$:
\begin{equation}
\begin{split}
T_1 = Tr \frac{L_P^2}{L^2} \left(\dot {\widetilde{Q}}_B^2\right) = Tr \frac{L_P^2}{L^2} \left(  \dot{R}_B \exp [i\theta_B] \dot{R}_B \exp[i\theta_B] + \dot{R}_B \exp [i\theta_B] R_B \exp[i\theta_B]i\dot{\theta}_B \right. \\ \left. +  R_B \exp [i\theta_B]i\dot{\theta}_B \dot{R}_B \exp[i\theta_B] + R_B \exp [i\theta_B] i\dot{\theta}_B R_B \exp[i\theta_B ] i\dot{\theta}_B \right)
\end{split}
\end{equation}
\begin{equation}
\begin{split}
T_2 = Tr \frac{L_P^4}{L^4} \left[\dot {\widetilde{Q}}_B\beta_2\dot{\widetilde {Q}}_F\right] = Tr \frac{L_P^4}{L^4}   \left(  \dot{R}_B \exp [i\theta_B]\beta_2 \dot{R}_F \exp[i\theta_F] + \dot{R}_B \exp [i\theta_B] \beta_2 R_F \exp[i\theta_F]i\eta\dot{\theta}_F \right. \\ \left. +  R_B \exp [i\theta_B]i\dot{\theta}_B\beta_2 \dot{R}_F \exp[i\theta_F] + R_B \exp [i\theta_B] i\dot{\theta}_B \beta_2 R_F \exp[i\theta_F ] i\eta\dot{\theta}_F \right)
\end{split}
\end{equation}
\begin{equation}
\begin{split}
T_3 = Tr \frac{L_P^4}{L^4} \left[ \beta_1\dot{\widetilde{Q}}_F\dot{\widetilde {Q}}_B \right] = Tr \frac{L_P^4}{L^4} 
\left(  \beta_1\dot{R}_F \exp [i\theta_F] \dot{R}_B \exp[i\theta_B] + \beta_1\dot{R}_F \exp [i\theta_F]  R_B \exp[i\theta_B]i\dot{\theta}_B \right. \\ \left. +  \beta_1 R_F \exp [i\theta_F]i\eta\dot{\theta}_F \dot{R}_B \exp[i\theta_B] + \beta_1 R_F \exp [i\theta_F] i\eta\dot{\theta}_F  R_B \exp[i\theta_B ] i\dot{\theta}_B \right)
\end{split}
\end{equation}
\begin{equation}
\begin{split}
T_4 = Tr \frac{L_P^6}{L^6} \left[ \beta_1\dot{\widetilde{Q}}_F\beta_2\dot{\widetilde {Q}}_F\right] = 
\left(  \beta_1\dot{R}_F \exp [i\theta_F] \beta_2 \dot{R}_F \exp[i\theta_F] + \beta_1\dot{R}_F \exp [i\theta_F]\beta_2  R_F\exp[i\theta_F]i\eta\dot{\theta}_F \right. \\ \left. +  \beta_1 R_F \exp [i\theta_F]i\eta\dot{\theta}_F \beta_2\dot{R}_F \exp[i\theta_F] + \beta_1 R_F \exp [i\theta_F] i\eta\dot{\theta}_F \beta_2 R_F \exp[i\theta_F ] i\eta\dot{\theta}_F \right)
\end{split}
\end{equation}
The momenta can be worked out by taking appropriate trace derivatives of the Lagrangian, using the rules of differentiation from trace dynamics. The varied matrix should be moved to the extreme right by cyclic permutation inside the trace, keeping in mind that exchange of two fermionic matrices results in a change of sign in the overall expression.

The fermionic spin angular momentum is
\begin{equation}
\begin{split}
p_{F\theta} = \frac{\delta {\cal L}}{ \delta \dot{\theta}_F} =\frac{L_P^4}{L^4} \bigg[- R_B \exp[i\theta_B]i\dot{\theta}_B (\beta_1+\beta_2)R_F\exp[i\eta\theta_F]\eta + i\left(\dot{R}_B \exp[i\theta_B]\beta_1+\beta_2)R_F\exp[i\eta\theta_F]\eta\right)\bigg] \\ + {\cal O}\left(\frac{L_P^6}{L^6}\right)
\end{split}
\label{fs}
\end{equation}
The higher order terms do not contribute to the present discussion. As we will see below, this expression has the desired form for matching with the conventional discussion of spin in the Dirac equation. We note that this spin angular momentum is not a conserved quantity; the Lagrangian explicitly depends on $\theta_F$. 

The bosonic spin angular momentum is given by
\begin{equation}
\begin{split}
p_{B\theta} = \frac{\delta {\cal L}} {\delta \dot{\theta}_B} = -2 \frac{L_P^2}{L^2} R_B \exp [i\theta_B] \dot{\theta}_B R_B \exp[i\theta_B]
 +2i \frac{L_P^2}{L^2} \dot{R}_B \exp[i\theta_B] R_B \exp[i\theta_B] \\ + \frac{L_P^4}{L^4} (\beta_1 + \beta_2) \dot{R}_F\exp[i\eta\theta_F] R_B \exp [i\theta_B] +i\frac{L_P^4}{L^4} (\beta_1 + \beta_2) R_F \exp[i\eta\theta_F] \eta \dot{\theta}_F R_B \exp[i\theta_B]
\end{split}
\end{equation}

The fermionic and bosonic linear momenta are given by
\begin{equation}
\begin{split}
p_{BR} = \frac{\delta {\cal L}}{ \delta \dot{R}_B} =2\frac{L_P^2}{L^2} \exp [i\theta_B] \dot{R}_B  \exp[i\theta_B] + 
2i\frac{L_P^4}{L^4} \exp [i\theta_B] {R}_B  \exp[i\theta_B]\dot{\theta}_B \\
+\frac{L_P^4}{L^4} \exp[i\theta_B] (\beta_1 + \beta_2) \dot{R}_F \exp[i\theta_F] + i\frac{L_P^4}{L^4} \exp[i\theta_B](\beta_1+\beta_2)R_F\exp[i\eta\theta_F]\eta\dot{\theta}_F
\end{split}
\end{equation}
\begin{equation}
\begin{split}
p_{FR} = \frac{\delta {\cal L}}{ \delta \dot{R}_F} =\frac{L_P^4}{L^4}\left[ \exp[i\eta\theta_F]\dot{R}_B \exp[i\theta_B] (\beta_1 +\beta_2) +
i\exp[i\eta\theta_F] R_B \exp[i\theta_B] \dot{\theta}_B (\beta_1 +\beta_2) \right] \\ + {\cal O}\left(\frac{L_P^6}{L^6}\right)
\end{split}
\end{equation}

In our earlier work, we constructed the variables $\widetilde{Q}_B$ and $\widetilde{Q}_F$ as follows:
\begin{equation}
{\dot{\widetilde{Q}}_B} = \frac{1}{L} (i\alpha q_B + L \dot{q}_B); \qquad  {\dot{\widetilde{Q}}_F} = \frac{1}{L} (i\alpha q_F + L \dot{q}_F)
\label{lilq}
\end{equation}
Here, the self-adjoint operator $q_B$ corresponds to the Yang-Mills potential in the classical limit, and $\dot{q}_B$ to gravity. $\dot{q}_F$ is the matter source for gravity, and $q_F$ is the current that sources the Yang-Mills fields. $\alpha$ is the gauge coupling constant. These operators are related to the standard Dirac operator $D_B$, and we also defined an operator $D_F$:
\begin{equation}
D_B \equiv \frac{1}{Lc}\;  \frac{dq_B}{d\tau}; \qquad   D_F \equiv \frac{L_P^2}{L^2}\frac{\beta_1+\beta_2}{2Lc}\;  \frac{dq_F}{d\tau}
\end{equation}
$D_B$ is defined such that in the commutative c-number limit where space-time emerges, it becomes the standard Dirac operator on a Riemannian manifold. $D_F$ is defined such that upon spontaneous localisation,  it gives rise to the classical action for a relativistic point particle. The modified Dirac operators which take into account the presence of the Yang-Mills potential $q_B$ and the corresponding current $q_F$, are given by
\[
D_{Bnewi} = \dfrac{1}{L} \dot{\widetilde{Q}}_{B} \qquad and \qquad D_{Fnewi} = \frac{L_P^2}{L^2}\frac{\beta_1+\beta_2}{2Lc} \dot{\widetilde{Q}}_{F}
\label{eq:ddirac}
\]
$q_B$ is related to the gauge-potential by $\alpha q_B/L^2 = \gamma^\mu A_\mu$, and $q_F$ is related to the gauge current.

The constancy of the bosonic momentum corresponding to $\widetilde{Q}_B$ implies that  we have a constant net Dirac operator which can be expressed as an eigenvalue equation given by:
\[
\left[D_{Bnewi} + D_{Fnewi}\right] \psi = \lambda \psi
\]
where the eigenvalues $\lambda$ are assumed to be $\mathbb{C}$-numbers [since the operator is bosonic]  and are independent of the Connes' time $\tau$.

We now  work out what this Dirac equation looks like in terms of the variables $\theta_B, \theta_F, R_B$ and $R_F$ introduced in the present paper. This will unearth the presence of spin as defined in the current paper, and show its presence in the conventional Dirac equation. By comparing the real and imaginary parts of Eqns. (\ref{qexp2}) and (\ref{lilq}) we obtain that
\begin{equation}
\dot{q}_B = \dot{R}_B(\cos\theta_B)   - R_B (\sin\theta_B) \dot{\theta}_B 
\end{equation} 
\begin{equation}
\frac{\alpha}{L} q_B = \dot{R}_B (\sin\theta_B) + R_B (\cos\theta_B) \dot{\theta}_B 
\end{equation}
If we make the assumption that introducing the gauge-potential does not change the background space-time geometry too much, we should have that $\theta_B$ is small and that $q_B$ is nearly the same as $R_B$ and $\dot{q}_B$ is nearly the same as $\dot{R}_B$. This trivialises the first of these equations [$q_B=q_B]$ whereas the second equation gives that ${\alpha} q_B/L^2= R_B\dot\theta_B /L$. The left hand side in this equality is of course the same as the contribution of the gauge potential to the Dirac operator $D_{Bnewi}$. We now show that  the right hand side is proportional to the spin angular momentum, as defined above. Let us look at the self-adjoint part of the expression (\ref{fs}) for fermionic spin $p_{F\theta}$. For small $\theta_B$ we can approximate it as
$-R_B \dot{\theta}_B (\beta_1 + \beta_2) R_F \exp [i\eta\theta_F] \eta$. This means that $p_{F\theta } \sim -(\alpha q_B/L) \widetilde{Q}_F\eta$ which represents the coupling of the fermion to the gauge potential. Equivalently, the term $\alpha q_B/L$ represents the correction to the Dirac operator enforced by the spin angular momentum $p_{F\theta}$. We also know that this correction term in the standard Dirac equation allows us to conclude that the electron has a gyromagnetic ratio of 2 and hence has a spin $\hbar/2$. We therefore conclude that our definition of the spin angular momentum $p_{F\theta}$ is consistent with the conventional understanding of spin in relativistic quantum mechanics. Of course this Dirac equation is still not on space-time, and evolution is with respect to Connes time; however the transition from here to the conventional Dirac equation on a space-time is straightforward [after spontaneous localisation of macroscopic systems gives rise to emergent space-time] - the Dirac operator is the standard one, and the gauge potential represents an external potential to which the fermion is coupled.

One subtle point to note is the following. In the definition of $\dot{\widetilde{Q}}_B$ and $\dot{\widetilde{Q}}_F$, the gravity part $\dot{q}_B$ and the gauge part $q_B$ have a relative $i$ factor between them, and the same is true for the pair $(\dot{q}_F, q_F)$. This is different from the standard theory where there is of course no $i$ factor between $D_B$ and the gauge potential $\alpha A$ in the modified Dirac operator $D_B + \alpha A$. We believe our construction to be more fundamental, as it allows us to think of the gauge interaction as the phase part of the complex $\widetilde{Q}_B$. 
This will also motivate that the aikyon lives in an octonionic space, as we will see in the discussion below.
Equivalence with the standard theory can be restored if  in the classical (gauge field plus current) Lagrangian we introduce an $i$ factor in the current part of the interaction term $j_\mu A^\mu$ and also an $i$ factor in front of the potential term $A^\mu$ in this interaction term, and make an overall change of sign for this term in the Lagrangian. And similarly, an overall change in  sign in front of the gauge-field Lagrangian, because $i^2 = -1$. In our approach, we were compelled to work with an imaginary gauge potential, because otherwise the matrix dynamics gives exponentially growing and decaying solutions, in Connes time, which seems undesirable. Imaginary potential enables the more reasonable oscillatory solutions [see Eqns. (35) and (36) of \cite{MPSingh}). The imaginary potential then strongly indicates that this could be the right Lagrangian for unifying gravity and the standard model. Furthermore, the imaginary potential allows for an anti-self-adjoint component to the Hamiltonian, enabling spontaneous localisation and recovery of the classical limit.

\section{Understanding Spin}
From the Lagrangian above, we can write the first integrals for the equations of motion for $\widetilde{Q}_B$ and $\widetilde{Q}_F$. The two corresponding canonical momenta are constants of motion, implying that
\[
\widetilde{p}_{B} = \dfrac{\partial \mathcal{L}}{\partial \dot{\widetilde{Q}}_{B}} = \dfrac{L_{p}^{2}}{L^{2}} \biggl[\biggr. 2 \dot{\widetilde{Q}}_{B} + \dfrac{L_{p}^{2}}{L^{2}} (\beta_{1} + \beta_{2}) \dot{\widetilde{Q}}_{F} \biggl.\biggr]
\]
\[
\widetilde{p}_{F} = \dfrac{\partial \mathcal{L}}{\partial \dot{\widetilde{Q}}_{F}} = \dfrac{L_{p}^{4}}{L^{4}} \biggl[\biggr. \dot{\widetilde{Q}}_{B} (\beta_{1} + \beta_{2}) + \dfrac{L_{p}^{2}}{L^{2}} \beta_{1} \dot{\widetilde{Q}}_{F} \beta_{2} + \dfrac{L_{p}^{2}}{L^{2}} \beta_{2} \dot{\widetilde{Q}}_{F} \beta_{1} \biggl.\biggr]
\]
The presence of spin in quantum mechanics is also indicated from this expression above for the bosonic momentum, from which
the Dirac equation is constructed. It depends not only on the bosonic velocity but also on the fermionic velocity, which is related to the fermioinic spin angular momentum.

Here the conjugate momenta, $\widetilde{p}_{B}$ and $\widetilde{p}_{F}$ are constants as the trace Lagrangian is independent of $\widetilde{Q}_{B}$ and $\widetilde{Q}_{F}$, similar to what happened for pure gravity. This implies,
\[
2 \dot{\widetilde{Q}}_{B} + \dfrac{L_{p}^{2}}{L^{2}} (\beta_{1} + \beta_{2}) \dot{\widetilde{Q}}_{F} = C_{1}
\label{eq:dirac1}
\]
\[
\dot{\widetilde{Q}}_{B} (\beta_{1} + \beta_{2}) + \dfrac{L_{p}^{2}}{L^{2}} \beta_{1} \dot{\widetilde{Q}}_{F} \beta_{2} + \dfrac{L_{p}^{2}}{L^{2}} \beta_{2} \dot{\widetilde{Q}}_{F} \beta_{1} = C_{2}
\]
for some $C_{1}$ and $C_{2}$ which are constant bosonic and fermionic matrices respectively. We can deduce from  the definition of $\dot{\widetilde{Q}}_B$ and $\dot{\widetilde{Q}}_F$ in terms of $(q_B, \dot{q}_B, q_F, \dot{q}_F)$ as to how the former set of variables evolve. The latter set evolve as harmonic oscillators \cite{MPSingh}:
\[
q_{B} = B_{+} e^{i \, (\alpha\tau/L)} + B_{-} e^{-i \, (\alpha\tau/L)};
\qquad q_{F} = F_{+} e^{i \, (\alpha\tau/L)} + F_{-} e^{-i \, (\alpha\tau/L)}
\]
This implies that in the complex `plane' formed by $\dot{q}_B$ along the horizontal axis, and $iq_B$ along the vertical axis, the complex dynamical variable $\dot{\widetilde{Q}}_B$ executes  periodic motion with a time-period $L/\alpha$. The angular momentum associated with this periodic motion is the bosonic spin angular momentum. An analogous interpretation holds for the fermionic spin, vis a vis the motion of $\dot{\widetilde{Q}}_F$ in the complex plane formed by $\dot{q}_F$ and $iq_F$. In the emergent quantum theory, this spin angular momentum is quantised in units of Planck's constant, just like angular momentum is, except that, because fermions are described by odd-grade Grassmann matrices, their spin is half-integral.

To put it more physically, spin is the angular momentum associated with the motion of an aikyon in the Hilbert space of matrix dynamics. The motion takes place in the two dimensional `plane' formed by the self-adjoint and anti-self-adjoint parts of the Grassmann matrix which describes an aikyon. The self-adjoint part relates to gravity and the anti-self-adjoint part to Yang-Mills gauge interactions. We can decompose this motion into a sum of  linear motion and angular motion. Since in both the linear motion as well as in angular motion, both the self-adjoint and anti-self-adjoint parts vary, each of these motions relate both to gravity and to gauge fields. However, since spin gets switched on only after the imaginary axis of the plane is switched on because of introducing gauge fields, it could be the case that there is an intimate connection between spin and gauge interactions. In particular, since spin relates to torsion in geometry, one should investigate if gauge interactions are manifestations of torsion, and of a complex anti-symmetric part to the space-time metric. This kind of a suggested unification of gravity and gauge fields on a complex plane might help get rid of the need for extra hidden space-time dimensions as required in Kaluza-Klein theories. The fact that gauge-interactions are related to the phase which obeys periodic boundary conditions might help understand why the standard model symmetry groups have to do with rotational invariance, whereas gravity, related to the amplitude $R_B$ has to do with diffeomorphisms. 

Another way to think of spin is to regard the self-adjoint fermionic position and velocity operators, $\dot{q}_F$ and $q_F$, as the real and imaginary parts of the complex velocity $\dot{\widetilde{Q}}_F$ [which is what they precisely are]. Spin is the momentum associated with change of phase during evolution (in Connes time) in the complex plane defined by position and velocity. Linear momentum is associated with change in amplitude of $\dot{\widetilde{Q}}_F$ during evolution.

\section{Octonions, trace dynamics, and non-commutative geometry: a case for unification in spontaneous quantum gravity}

Let us once again write down our fundamental Lagrangian for the aikyon, as given in Eqn. (62) of \cite{MPSingh}, and above in (\ref{eq:tracelagn}). This describes the unification of gravity with Yang-Mills and fermions with gravity:
\[
 \mathcal{L} = Tr \biggl[\biggr. \dfrac{L_{p}^{2}}{L^{2}} \biggl(\biggr. \dot{\widetilde{Q}}_{B} + \dfrac{L_{p}^{2}}{L^{2}} \beta_{1} \dot{\widetilde{Q}}_{F} \biggl.\biggr) \biggl(\biggr. \dot{\widetilde{Q}}_{B} + \dfrac{L_{p}^{2}}{L^{2}} \beta_{2} \dot{\widetilde{Q}}_{F} \biggl.\biggr) \biggl.\biggr]
\label{eq:tracelagn2}
\]
We rewrite as before,  self-adjoint bosonic operators $R_B$ and $\theta_B$, and self-adjoint fermionic operators $R_F$ and $\theta_F$, as follows:
\begin{equation}
\widetilde{Q}_B \equiv R_B\; \exp i\theta_B \ ;   \qquad \widetilde{Q}_F \equiv R_F \; \exp i\eta \theta_F
\label{que22}
\end{equation}
Each of these four newly introduced  self-adjoint operators are functions of Connes time, and are the four configuration variables which define the aikyon.  Figs. 1 and 2 below attempt to give a visualisation of the aikyon.
\begin{figure}[!htb]
        \center{\includegraphics[width=\textwidth]
        {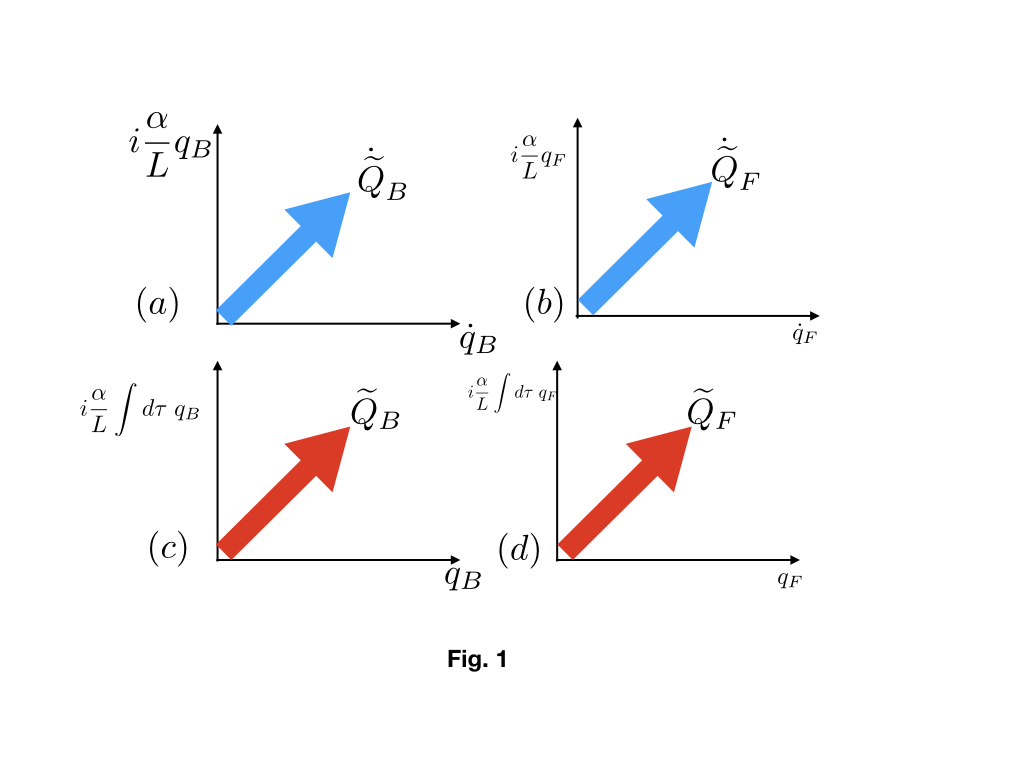}}
        \caption{ A cartoon for visualising the aikyon. Figs. 1a and 1b depict the aikyon velocity, as expressed in Eqn. (\ref{lilqu}). Figs. 1c and 1d show the aikyon after the integration of  the velocity.}
             \end{figure}
\begin{figure}[!htb]
        \center{\includegraphics[width=\textwidth]
        {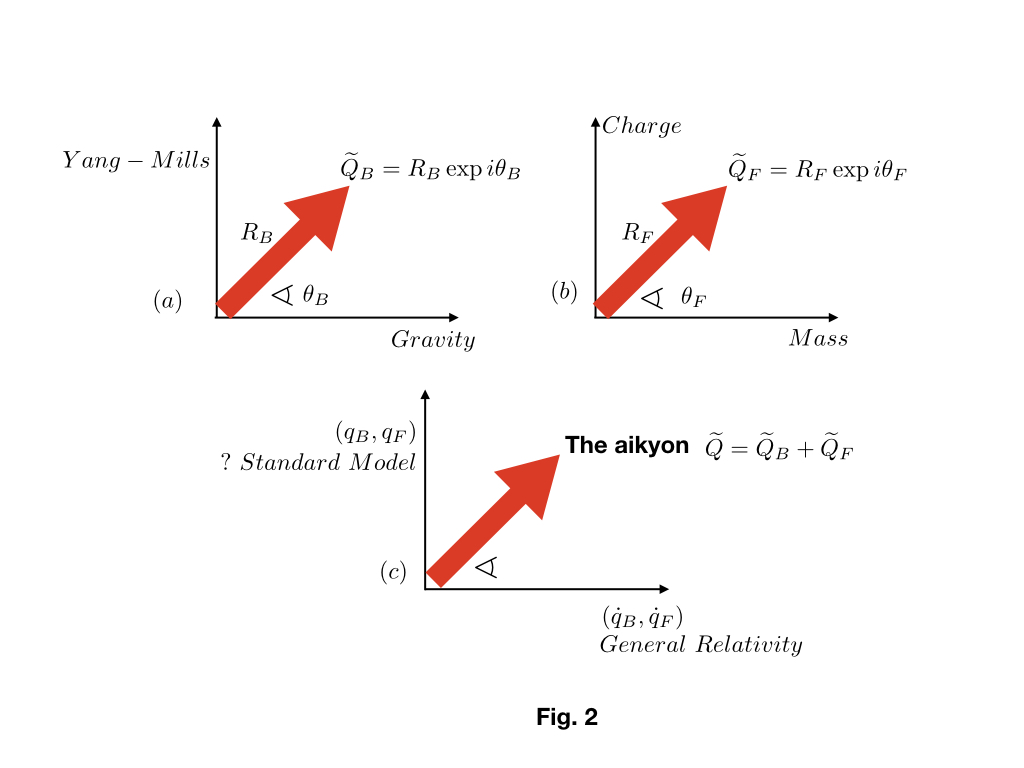}}
        \caption{ A second cartoon for visualising the aikyon. Figs. 2a and 2b depict the aikyon amplitude and phase, as expressed in Eqn. (\ref{que}). Fig. 2c  symbolically shows the aikyon in octonionic space: the horizontal axis represents would be spacetime in terms of the four quaternionic directions and the vertical axis represents the other four octonionic directions.}
             \end{figure}

Since the Lagrangian of the theory is invariant under global unitary transformations, and since it describes a unification of gravity, Yang-Mills fields and Dirac fermions, it is pertinent to ask if the symmetry group can be so chosen as to unify gravity with the standard model. There are hints in our analysis and in existing work in the literature, that the answer could be yes. In the spirit of opening up our proposal for further investigation, we motivate below, tentatively,  that this symmetry group could be $G_2$, which is the smallest of the exceptional Lie groups. $G_2$ is also the group of automorphisms of the eight dimensional space of octonions. There seems to be a strong possibility that the aikyon lives in an octonionic space.

Why do we say so? The major hint comes from the work of Cohl Furey \cite{f1, f2, f3} (as well as many others, see e.g. \cite{Gillard} and references therein), who building on earlier work \cite{gursey, Okubo, Dixon, Marques1, Marques2}, has shown that the Dixon algebra - the product of the four division algebras [reals, complex numbers, quaternions and octonions] - can explain within itself several of the features of the standard model. This algebra naturally splits into two parts, the algebra of complex quaternions and the algebra of complex octonions.

Let us first focus on the complex quaternions. As is well-known,  from this algebra one can obtain the Clifford algebra $Cl(2)$, which implies the Lorentz algebra in a four-dimensional space-time. It is this Lorentz symmetry which is of interest to us. Let us look at the Lagrangian in Eqn. (1) above, which describes the coupling of gravity to fermions in our matrix dynamics at the Planck scale. We have so far in our work not talked of any coordinate space in which the aikyon lives. With a view to recovering Lorentz invariance and causal structure of the 4-d classical space-time at low energies, we now propose that the aikyon $q=q_B + q_F$ carries a quaternionic index $\mu$ which runs from one to four, and this index runs over the usual quaternionic  $(1, i, j, k)$. We recall that the three complex indices obey the cyclic multiplication rule, and also anti-commute amongst themselves. Thus, in this Lagrangian, each of the complex Grassmann-valued matrices $\dot{q}_B$ and $\dot{q}_F$ is assumed to be written in components, as a sum of four components over the four quaternionic coordinates. These are hence Grassmann valued matrices over the complex quaternions. Their symmetry group is hence the Lorentz group, as follows from the preceding discussion. We must recall of course that these matrices do not commute with each other, and hence we are not talking of classical space-time nor classical causal 
light-cone structure. Such structure - 4-d classical space-time and causality - emerges though, when spontaneous localisation localises the fermionic part of the aikyon to one specific eigenvalue of $q_F$. Local Lorentz symmetry is preserved during this localisation. The curvature of the emergent space-time arises because of an effective gauging of the Lorentz group, because the spontaneous collapse of different aikyons happens to different eigenvalues of the Dirac operator. In terms of Fig. 1 and Fig. 2 this pure gravity aikyon always stays along the horizontal axis [no Yang-Mills fields]. Hence, the horizontal axis represents the four quaternionic directions.

So much for the pure gravity plus fermion Lagrangian of Eqn. (1). Let us now consider the case when Yang-Mills fields are brought in, as described by the Lagrangian in Eqns. (\ref{laym}) and (\ref{eq:tracelagn}). Can one still stay within the quaternionic space? The answer is a definite no, and one has to go the octonionic space, as we now argue. Let us denote the terms in the two big  brackets in the Lagrangian (\ref{laym}) as $T_1$ and $T_2$ respectively, and expand the terms inside each of the brackets, and write them explicitly, combining the terms as follows (ignoring to exhibit the trace and the $L_P^4 / L^4$ factor):
\begin{equation}
T_1 = (L\dot{q}_B + i\alpha q_B)  + \frac{L_P^2}{L^2} \beta_1 (L\dot{q}_F +i\alpha q_F)
\end{equation}
\begin{equation}
T_2 = (L\dot{q}_B + i\alpha q_B)  + \frac{L_P^2}{L^2} \beta_2 (L\dot{q}_F +i\alpha q_F)
\end{equation}
The introduction of the Yang-Mills field $q_B$ amounts to a complex rotation of the self-adjoint $\dot{q}_B$. This certainly takes us out of the quaternionic space of gravity, and is a lift from the horizontal axis into the complex plane. And since $q_{B}$ has as many components as $\dot{q}_B$, we have four new directions to be denoted by the vertical axis. It appears natural then to identify the eight directions as an octonionic space. The horizontal axis is now assumed to represent four of the octonionic directions: the real one, and any one of the quaternionic triples from the Fano plane. These four constitute the would-be space-time, and the Yang-Mills fields lie on the four vertical directions: this harmonious split of gravity and gauge fields into equal number of directions [four each] is elegant, undoubtedly. We think of the octonionic space as the sum of two quaternionic spaces, one of which is space-time, and the other is the internal space for the standard model interactions.

Another way to see that gauge fields indicate doubling of space dimensions is to recall that in quantum mechanics, the gauge potential is included by modifying the Dirac operator $D_B$ to $(D_B - eA)$: the bosonic momentum operator is modified by adding a configuration variable, {\it not} by adding another momentum variable. This is quite strange in itself, even though we know that it is the right thing to do. From the preceding discussion it is apparent that this results from gravity being velocity aspect of the aikyon, and Yang-Mills being the position aspect. Being independent, they must be represented on different directions in the space in which the aikyon moves. We may symbolically write the Lagrangian as $Tr [D^2]$, where $D$ is the generalised Dirac operator, and this then looks like the `energy' of a free particle, and we are asking what is the group of transformations in the octonionic space which leaves this expression for the energy unchanged. This seems to capture the two symmetries, Lorentz invariance and gauge-invariance, in one common unified description.

The introduction of Yang-Mills fields compels us to deal with octonions and the full Dixon division algebra. The same argument holds for the fermionic part of $T_1$ and $T_2$. Note that we cannot discuss the Yang-Mills part just on the vertical axis, because the Lagrangian involves gravity also! We cannot drop the gravity part and write a Lagrangian only in terms of $q_B$ and $q_F$. We have to work on the full plane and hence deal with complex octonions. Bringing in Yang-Mills forces on us the unification with gravity. Moreover, we can now nicely understand spin as resulting from rotation in the octonionic space. Spin remains mysterious if we try to understand it from the quaternionic subspace that is space-time.

Working with the algebra of complex octonions Furey showed that the Clifford algebra $Cl(6)$ is obtained. It is known that $Cl(6)$ gives a highly elegant construction of one generation of the standard model, as shown by various researchers \cite{Stoica, Chisholm, Trayling}. Of course one could ask why octonions are needed, if $Cl(6)$ already does the job. One can say that complex quaternions naturally imply $Cl(6)$, rather than having to pick it arbitrarily. More importantly, from our point of view, octonions are absolutely essential for the unification of gravity and the standard model: complex quaternions imply the Lorentz symmetry and gravity, and complex octonions imply one generation of the standard model.

In what appears to be a significant development, Gillard and Gresnigt \cite{Gillard} have recently extended the very elegant work of Stoica \cite{Stoica} and proposed that when spin degrees of freedom are included along with $Cl(6)$ via $Cl(2)$, the Clifford algebra describing the standard model is extended to $Cl(8)$. This also extends all the results of a single generation to three generations, while still including the spin degrees of freedom. This would be a remarkable demonstration that three generations of the standard model can be unified with gravity, and already serves the purpose of the symmetry we were  seeking for our Lagrangian. To our understanding this utilises the full Dixon algebra, combining complex quaternions with complex octonions. The symmetries along the horizontal direction in Figs. 1 and 2 are described by $Cl(2)$, those along the vertical direction by $Cl(6)$, and those in the full plane by $Cl(8)$. This appears to complete Furey's programme of describing the unification of gravity and the standard model, using the four division algebras. 

What remains to be seen is if our matrix dynamics can predict the Higgs field and values of the standard model parameters. This investigation is in progress. The decoupling between the Planck scale matrix dynamics  and the low energy physics might come about naturally because in Adler's trace-dynamics quantum field theory arises below the Planck scale by coarse-graining over the ``fast'' Planck-scale modes of the matrix dynamics. The eigenvalue equation for the Hamiltonian of the matrix dynamics is analogous to a time-independent Schr\"odinger equation, and the eigenvalues might be able to predict values of the standard model parameters. This is currently being investigated.

In our work, local quantum field theory arises as a low energy limit of this Planck scale dynamics of the so-called aikyons, and classical general relativity arises as a consequence of spontaneous localisation of a large number of entangled aikyons. We propose that classical curved space-time and Yang-Mills fields arise from an effective gauging which results from the collection of symmetry groups of the spontaneously localised fermions. Our work suggests that we live in an eight-dimensional octonionic universe: four of these dimensions constitute space-time and the other four constitute the octonionic internal directions on which the standard model forces live.

There are other encouraging signs that our theory could be on the right track. The aikyon behaves like a two dimensional entity, because its Lagrangian involves two unequal constant matrices $\beta_1$ and $\beta_2$. This makes the aikyon very much like a 2-D string. Moreover, the aikyon evolves in the 8-D octonionic space in Connes time [this time then effectively serves as the ninth dimension] and it has been suggested \cite{Baez} that a string in 10-D spacetime evolves as if it resides in the 8-D octonionic space. The multiplicative chain of elements of the Dixon algebra has ten generators, nine of which act like spatial dimensions, whereas the tenth one acts like time. Also, in a very interesting work, Perelman \cite{Perelman}  has constructed a grand unified theory including gravity, based on the Dixon algebra. All these are encouraging pointers that the role of $G_2$ and $Cl(8)$ should be investigated further in our theory, for predicting standard model parameters from our theory.

Our theory bears resemblance to earlier studies of matrix models. In one class of such models non-commutative gravity is emergent from Yang-Mills matrix models (see e.g.the review by Steinacker \cite{Steinacker}). Our 8D theory is indeed a matrix model, from which classical space-time is emergent.   A strong possibility of connection with string theory / M-theory has emerged from further investigation of the present approach \cite{Singh2020DA}. The key new points are the following. The present approach amounts to redoing string theory by replacing the laws of quantum field theory by those of trace dynamics, at the Planck scale. Quantum field theory is emergent  from trace dynamics, at lower energies. Further, evolution takes place in Connes time, in an octonionic space. The aikyon is a 2-brane evolving in this 8D non-commutative space. Thus in totality, this is an eleven dimensional theory [8+2+1] and is reminiscent of M-theory. It has been noted by Baez and Huerta \cite{Baez} that string theory in 10 dimensions plus an extra time dimension is equivalent to a string evolving in octonionic space. Furthermore, the Hamiltonian in our theory is in general not self-adjoint at the Planck scale: this allows for dynamical compactification of the extra dimensions through spontaneous localisation. Whereas quantum systems continue to live in the eight octonionic dimensions. [The emergent quantum theory Hamiltonian is self-adjoint]. The thickness of these extra dimensions is not universal. It will vary from aikyon to aikyon, and is of the order of the Compton wavelength of the quantum particle. The connection with string theory deserves to be explored further: the point being that our theory also investigates a string / 2-brane in higher dimensions, but uses trace dynamics instead of quantum field theory, brings in Connes time, and allows for a non-self-adjoint Hamiltonian, These points of departure appear to restore predictability in string theory, and provide a clear connection with the standard model. The great advantage of using trace dynamics at the Planck scale is that the Lagrangian of the theory does not have to be quantised. The theory is pre-quantum, and quantum theory is emergent from it.

\bigskip

\noindent{\bf Acknowledgements:} I would like to thank Basudeb Dasgupta, Anmol Sahu and Cristi Stoica for helpful discussions.

\vskip 0.4 in
\centerline{\bf REFERENCES}
\bibliographystyle{unsrt}
\bibliography{biblioqmtstorsion}

\end{document}